\documentclass{aa}  

\usepackage{placeins}
\usepackage[hidelinks]{hyperref}
\usepackage[dvipsnames]{xcolor}
\usepackage{tabularx}
\hypersetup{colorlinks, linkcolor={blue!70!black}, citecolor={blue!70!black}, urlcolor={blue!70!black}}
\usepackage{graphicx}
\usepackage{txfonts}
\begin{document}

   \title{Chronology of our Galaxy from \textit{\textbf{Gaia}} colour–magnitude diagram fitting (ChronoGal)}

   \subtitle{IV. On the inner Milky Way stellar age distribution}

   \author{Tomás Ruiz-Lara\inst{1,2}
          \and David Mirabal\inst{3,4}
          \and Carme Gallart\inst{3,4}
          \and Robert Grand \inst{5}
          \and Francesca Fragkoudi \inst{6}
          \and Isabel Pérez \inst{1,2}
          \and Santi Cassisi\inst{7,8}
          \and Emma Fernández-Alvar \inst{3,4}
          \and Anna B. Queiroz\inst{3,4}
          \and Guillem Aznar-Menargues\inst{3}
          \and Yllari K. González-Koda \inst{1,2}
          \and Alicia Rivero \inst{3,4}
          \and Francisco Surot \inst{3,4}
          \and Guillaume F. Thomas\inst{3,4}
          \and Rebekka Bieri\inst{9}
          \and Facundo A. Gomez\inst{10}
          \and R\"udiger Pakmor \inst{11}
          \and Freeke van de Voort \inst{12}
          }
   \institute{Universidad de Granada, Departamento de Física Teórica y del Cosmos, Campus Fuente Nueva, Edificio Mecenas, 18071 Granada, Spain \email{ruizlara@ugr.es}
             \and  
             Instituto Carlos I de Física Teórica y Computacional, Facultad de Ciencias, E-18071 Granada, Spain
             \and  
             Instituto de Astrofísica de Canarias, Calle Vía Láctea s/n, E-38206 La Laguna, Tenerife, Spain
             \and  
             Departamento de Astrofísica, Universidad de La Laguna, 38205 La Laguna, Tenerife, Spain
             \and
             Astrophysics Research Institute, Liverpool John Moores University, 146 Brownlow Hill, Liverpool, L3 5RF, UK
             \and
             Institute for Computational Cosmology, Department of Physics, Durham University, South Road, Durham, DH1 3LE, UK
             \and
             INAF-Osservatorio Astronomico d’Abruzzo, Via Mentore Maggini s.n.c., 64100 Teramo, Italy
             \and 
             INFN, Sezione di Pisa, Largo Pontecorvo 3, 56127 Pisa, Italy
             \and 
             Department of Astrophysics, University of Zurich, Zurich, Switzerland
             \and
             Departamento de Astronomía, Universidad de La Serena, Av. Juan Cisternas 1200 Norte, La Serena, Chile
             \and
             Max-Planck-Institute for Astrophysics, Karl-Schwartzschild-Strasse 1, 85748 Garching, Germany
             \and
             Cardiff Hub for Astrophysics Research and Technology, School of Physics and Astronomy, Cardiff University, Queen’s Buildings, Cardiff CF24 3AA, UK
             \\
             }
   \date{Received September 15, 1996; accepted March 16, 1997}
   
 
  \abstract{The Milky Way's inner region is dominated by a stellar bar and a boxy-peanut shaped bulge. However, which stellar populations inhabit the inner Galaxy or how star formation proceeded there is still unknown. The difficulty in studying these stars stems from their location in dense regions that are strongly impacted by extinction and crowding effects. In this work, we use star formation histories computed in the solar neighbourhood using Gaia Colour-Magnitude Diagram fitting to shed light onto the evolution of the central regions of our Galaxy. For that, we have obtained precise age distributions for the non-negligible amount of super metal-rich stars ([M/H] $\sim$ 0.5) in the solar neighbourhood (more than 5$\%$ of the total stars within 400 pc of the plane). Assuming that these stars were born in the inner Galaxy and migrated outwards, those distributions should be indicative of the true stellar age distribution in the inner Galaxy. Surprisingly, we find that these age distributions are not continuous but show clear signs of episodic star formation ($\sim$~13.5, 10.0, 7.0, 4.0, 2.0 and less than 1~Gyr ago). Interestingly, with the exception of the 4~Gyr event, the timings of the detected events coincide with the formation of the primitive Milky Way and with known merging events or satellite encounters (Gaia-Enceladus-Sausage, Sagittarius dwarf galaxy, and the Magellanic Clouds), suggesting that these could have induced enhanced and global star-forming episodes. These results are compatible with a scenario in which Gaia-Enceladus-Sausage is responsible for the formation of the bar 10 Gyr ago. However, we cannot associate any accretion counterpart with the 4-Gyr-ago event, leaving room for a late formation of the bar, as previously proposed. A qualitative comparison with the Auriga Superstars simulations also finds metal-rich stars in the solar neighbourhood-like regions, formed at discrete times and migrated from the inner parts of barred galaxies, suggesting a possible link to bar dynamics and satellite accretion. This novel analysis allows us to indirectly witness the evolution of the inner Milky Way and constrain dynamical models of the Milky Way bar.}



   \maketitle
%

\section{Introduction}

There is a clear consensus favouring the barred nature of our Galaxy \citep[e.g.][]{1994ApJ...429L..73S, mcwilliam_zoccali2010, saito2011, saito2012, anders2019}. In addition, detailed studies of the kinematics of stars populating the inner Galaxy show that a large fraction of stars follow a cylindrical rotation \citep[e.g.][]{sumi2003, howard2009, shen2010}, indicative of a secular, boxy-peanut bulge \citep[see also][]{dwek1995, wegg_gerhard2013}, linked to bar-related mechanisms such as the buckling instability (\citealt[][]{athanassoula2005}, see also \citealt{dimatteo2016, Fragkoudi2020bulge} comparing observations with simulations). Thus, all evidence indicates that we live in a barred Galaxy with a secularly evolved, boxy peanut-shaped bulge \citep[][]{kormendy2004}. However, given the observational limitations to study in detail these inner parts \citep[e.g.][]{2020A&A...644A.140S}, there is a great deal of uncertainty about the stellar populations inhabiting those areas.

In order to shed light onto the properties of the inner Milky Way (MW) stars, a myriad of photometric and spectroscopic surveys have targeted this area, mainly focusing on the bulge (e.g. VVV/VVVX, \citealt{minniti2010, 2024A&A...689A.148S}; OGLE IV, \citealt{udalski2015}; BRAVA, \citealt{rich2007}; ARGOS, \citealt{freeman2013}; GIBS, \citealt{zoccali2014}; APOGEE, \citealt{majewski2017}). The results from these works show that the inner Galaxy is much more complicated than previously anticipated. For instance, the prevailing view of the bulge being predominantly old is evolving as our understanding of the bulge stellar populations advances.

Early works computing ages for the bulge stars suggest that it is eminently old, composed of stars older than 10 Gyr \citep[e.g.][]{zoccali2003, clarkson2011, barbuy2018ARAA, renzini2018, bernard2018MNRAS}. However, other evidence indicates that, together with this old population, young and intermediate age stars are also present \citep[][]{vanloon2003, bensby2013, catchpole2016, bensby2017}. Especially interesting is the discontinuous age distribution found by \citet[][]{bensby2017}, with more than 35$\%$ of the stars analyzed younger than 8 Gyr, advocating for the presence of several episodes of star formation. On the other hand, \citet[][]{hasselquist2020}, despite finding some evidence favouring the existence of an intermediate-age stellar population (2 to 5 Gyr, mainly metal-rich stars), still conclude that most of the bulge populations should be older than 8 Gyr. Fortunately, there are strategies to evade the inherent difficulties of studying the inner Galaxy by focusing on the solar neighbourhood.

Galaxies, including our own, are living entities, with stars moving across them. In fact, the mixture of stellar ages and metallicities observed in the solar vicinity \citep[e.g.][]{carlberg1985, edvardsson1993, feltzing2001, bergemann2014} can only be explained if stellar radial migration is taken into account \citep[e.g.][]{2009MNRAS.398..591S, schonrich2009,  Roskar2012Radial, 2012MNRAS.425..969P, Halle2015churblur}. In particular, non-axisymmetric structures such as bars and spirals are natural redistributors of angular momentum, and thus, of stars from and to the inner Galaxy \citep[][]{MinchevFamaey2010, Minchev2018BirthRadii, 2024A&A...690A.147H}. As a consequence of this radial migration, stars originally born in the inner Galaxy can be found nowadays in the solar vicinity \citep[e.g.][]{hayden2015, Hayden2018churning, 2023A&A...669A..96D, Nepal2024_YoungBar} where they can be studied in more detail thanks to their closeness. Particularly, we can obtain photometric and spectroscopic data of main sequence and subgiant branch stars, which are key for age derivation. Therefore, the study of stars migrated from the inner Galaxy to the solar neighbourhood opens a new window to explore the Galactic inner parts.

In this work we use Colour-Magnitude Diagram (CMD) fitting applied to Gaia data \citep[][]{Prusti2016TheGaiaMission} to provide the age distribution of the most metal-rich stars in the solar vicinity. These stars are very likely to have been born near the center of the MW \citep[e.g.][]{MiglioChiappini2021_Kepler}, and thus, offer an alternative avenue to study the stellar age distribution of the inner Galaxy. We qualitatively compare the observational results with cosmological simulations from the Auriga Superstars suite\footnote{We note here that the bulge region of the original Auriga simulations were thoroughly studied in \citet[][]{gargiulo2019}, demonstrating a prevalence of pseudo-bulges.} \citep[][Fragkoudi et al. in prep.]{Grand2023, pakmor_superstars} to provide theoretical insights to our results. This paper is structured as follows. In Sect.~\ref{sec:data} we describe the subset of stars from Gaia that we analyse in this work. Section~\ref{sec:method} presents a brief overview on CMDft.Gaia, i.e. the methodology to extract Star Formation Histories (SFHs) from Gaia CMDs. The main results and discussion are provided in Sects.~\ref{sec:results} and~\ref{sec:discussion}. Conclusions are outlined in Sect.~\ref{sec:conclusions}.


\section{Gaia data}
\label{sec:data}

Precise astrometry and photometry from the Gaia mission \citep{Prusti2016TheGaiaMission, GaiaDR2_2018_Brown2018, GaiaDR3_2023Vallenari} is revolutionising our knowledge of the MW \citep[e.g.][]{helmi2018, belokurov2018_sausage, 2018A&A...618A..93C, haywood2018twoseq}. In particular, it allows for the derivation of age and metallicity distributions for large samples of stars from the analysis of CMDs in the absolute magnitude plane \citep[][]{Gallart2019NatAs, Ruiz-Lara2020Sgr, Gallart2024}. In this particular work, we use stars that are located within a cylinder of 1 kpc of radius centered at the Sun and at a maximum distance\footnote{Given the proximity of the sources and the quality of the Gaia parallaxes, the inverse of the parallax is a good approximation to the real distance in our case \citep[][]{BailerJones2018, luri2018}. In order to compute the distance to each star, we first correct its parallax using individual zero-point offsets \citep[][]{Lindegren2021_parallaxBias} from the {\tt gaiadr3-zeropoint} Python package. We also include a systematic uncertainty of 0.015 mas in the zero-point by adding this in quadrature with the parallaxes uncertainties.} of 3.5 kpc. 

To construct the CMDs in the absolute plane corresponding to this set of stars, we need to correct for the extinction from our own MW. For this, we use two different 3D extinction maps \citep[][]{Green2019, 2022A&A...661A.147L} to ensure that our results do not depend on this choice. We transform reddening measurements to the Gaia photometric system using the recipes described in \citet[][]{2019ApJ...886..108F}. Then, to compile a set of stars with the highest possible photometric quality we apply a series of quality cuts. First, we keep only those stars whose extinction in $G$-band (A$_G$) is below 0.5 magnitudes as largely extincted stars are likely to be affected by larger errors in the reddening determination, and thus, their position in the CMD is more uncertain. 

Then, we identify and remove stars with unreliable photometry based on \verb|phot_bp_rp_excess_factor|: 
\begin{equation*}
    \rm 0.001 + 0.039 \times bp\_rp < \log(phot\_bp\_rp\_excess\_factor)
\end{equation*}
 and 
 \begin{equation*}
     \rm \log(phot\_bp\_rp\_excess\_factor) < 0.12 + 0.039 \times bp\_rp
 \end{equation*}
  where \verb|bp_rp| is the observed G$_{\rm BP}$ - G$_{\rm RP}$ colour.
 
In the same line of keeping only those stars with a position in the CMD as precise as possible, we only retain stars with \verb|parallax_over_error| above 5 (i.e. ensuring a parallax relative uncertainty below 20$\%$). Of all quality cuts, the most restrictive one is the one related to the reddening (A$_G$ above 0.5), affecting mainly the volumes located closer to the disc (up to 47$\%$ stars removed by this cut using the \citealt[][]{2022A&A...661A.147L} dust map, slightly lower percentages using the \citealt[][]{Green2019} dust map). However,  this effect drastically decreases with height; by height 0.3 kpc, less than 15$\%$ of the stars are removed because of this. Thus, the results presented in this paper are not affected by this completeness limitation. The rest of quality cuts reduces only by 2$\%$ (at most) the number of stars.

This sample of stars is then divided geometrically in layers according to their vertical distance to the Galactic plane (above and below). In this way, the maximum height of 3.5 kpc is divided into 16 volumes using the following limits:
\begin{equation*}
\pm z \rm \: bins = [0.0, 0.05, 0.1, 0.15, 0.2, 0.3, 0.4, 0.5, 0.6, 0.7, 0.8, 0.9, 
\end{equation*}
\begin{equation*}
1.0, 1.2, 1.6, 2.3, 3.5] \: \rm kpc.
\end{equation*}
In this way, volume 3 will correspond, for instance, to stars with heights (in absolute values\footnote{We computed solutions for three different samples: above the plane, below the plane, and above and below together. The results were totally compatible, suggesting symmetry. For this reason, we decided to bin together stars with positive  and negative $z$-values, allowing us to have a larger number of stars per volume and, in consequence, better fits.}) from 0.1 to 0.15 kpc (i.e. above and below the plane).

Together with this main Gaia dataset (basically information from the \verb|gaia_source| catalogue), we also inspect the high-quality stellar chemophysical parameters from the Gaia DR3 GSP-Spec catalogue  \citep[][see Sect.~\ref{sec:recio}]{2023A&A...674A..38G}.

\section{Gaia colour-magnitude diagram fitting: CMDft.Gaia}
\label{sec:method}

This work is part of the ChronoGal project \citep[Chronology of our Galaxy from Gaia Colour-Magnitude Diagram-fitting,][]{Gallart2024}. ChronoGal is an ambitious project aiming at boosting our knowledge on how our Galaxy formed and evolved by providing one of the most sought combinations of parameters in Galactic Archaeology: precise stellar ages and metallicities. ChronoGal inherits the knowledge from decades of extracting SFHs of Local Group dwarf galaxies using CMD-fitting techniques\footnote{CMD-fitting techniques are based on the comparison of observed CMDs with synthetic model ones (based on stellar evolution theory) in order to derive the best combination of simple stellar populations that fits the observed CMD, i.e. recovering the age and metallicity characteristics of the stars in the analysed system.} \citep[][]{gallart1999, monelli2010tucana, monelli2010cetus, hidalgo2011, Gallart2015, Rusakov2021Fornax, Ruiz-Lara2021LeoI} to develop \textit{CMDft.Gaia}.  \textit{CMDft.Gaia} is an updated set of tools following the philosophy of previous works \citep[][]{apariciogallart2004, apariciohidalgo2009, bernard2018proc, Ruiz-Lara2021LeoI}, but tailored to Gaia data.

We use ChronoSynth to create a synthetic mother CMD composed of 120 million stars ($M_{G}$ brighter than 5) from the solar-scaled version of the BaSTI-IAC stellar evolution library \citep[][]{Hidalgo2018basti}. We include a 30$\%$ fraction of unresolved binaries ($\beta$), allowing for a minimum mass ratio for the binary members of 0.1 (q$_{\rm min}$). This synthetic population is created assuming a Kroupa initial mass function \citep[][]{kroupa1993}, and covers ages from 0.02 to 13.5 Gyrs, and global metallicities ([M/H]) from -2.2 to 0.45. The populations in this mother synthetic CMD cannot be directly compared to any observed Gaia CMD. Thus, we simulate in it completeness and uncertainties using DisPar-Gaia \citep[]{Ruiz-Lara2021LeoI, Ruizlara2022_HS, alvar2025}. For this, together with the samples described in Sect.~\ref{sec:data}, we defined auxiliary samples after the geometric definition but before applying any quality cuts \citep[full samples following the nomenclature introduced in][]{alvar2025}.  

The comparison between observed and model CMDs (created as combinations of simple stellar populations, and including completeness and uncertainty effects) is done using dirSFH. Following the extensive testing carried out in \citet[][]{Gallart2024},
we apply a shift of -0.035 and  0.040 mag to the colour and magnitude of the synthetic stars, respectively, in order to account for residual shortcoming in the bolometric corrections adopted to transfer the stellar models from the theoretical plane to the Gaia photometric system. We also use a weighted scheme for the fit based on the inverse of the variance of the stellar ages in each CMD pixel and the  "S" bins\footnote{"S" bins denotes the seeds in age and metallicity used to define single stellar populations in the synthetic CMD and are ages=[0.02, 0.06, 0.126, 0.192, 0.262, 0.334, 0.404, 0.469, 0.532,
0.596, 0.656, 0.718, 0.784, 0.857, 0.938, 1.028, 1.128, 1.
244, 1.391, 1.576, 1.81, 2.066, 2.337, 2.609, 2.882, 3.156, 3.427,
3.695, 3.978, 4.272, 4.581, 4.946, 5.389, 5.858, 6.351, 6.8
61, 7.372, 7.883, 8.393, 8.904, 9.415, 9.925, 10.436, 10.947, 11.457,
11.968, 12.479, 12.989, 13.5] Gyr. The weighing scheme refers to the importance that each region in the colour-magnitude plane has in the final fit.} \citep[see][for more information]{Gallart2024}. 

\section{Results}
\label{sec:results}

\begin{figure*}
\centering
\includegraphics[width=0.32\textwidth]{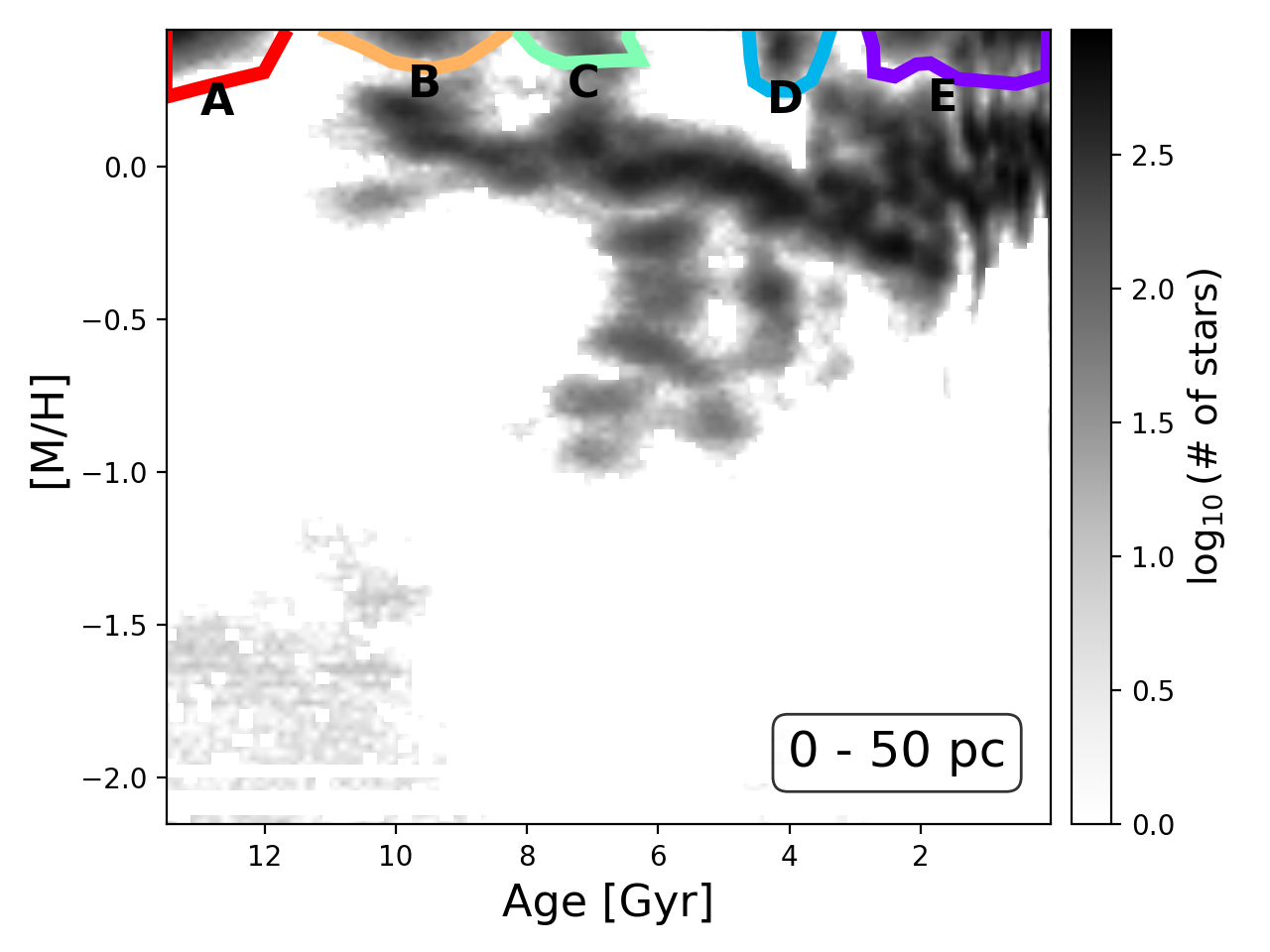} ~
\includegraphics[width=0.32\textwidth]{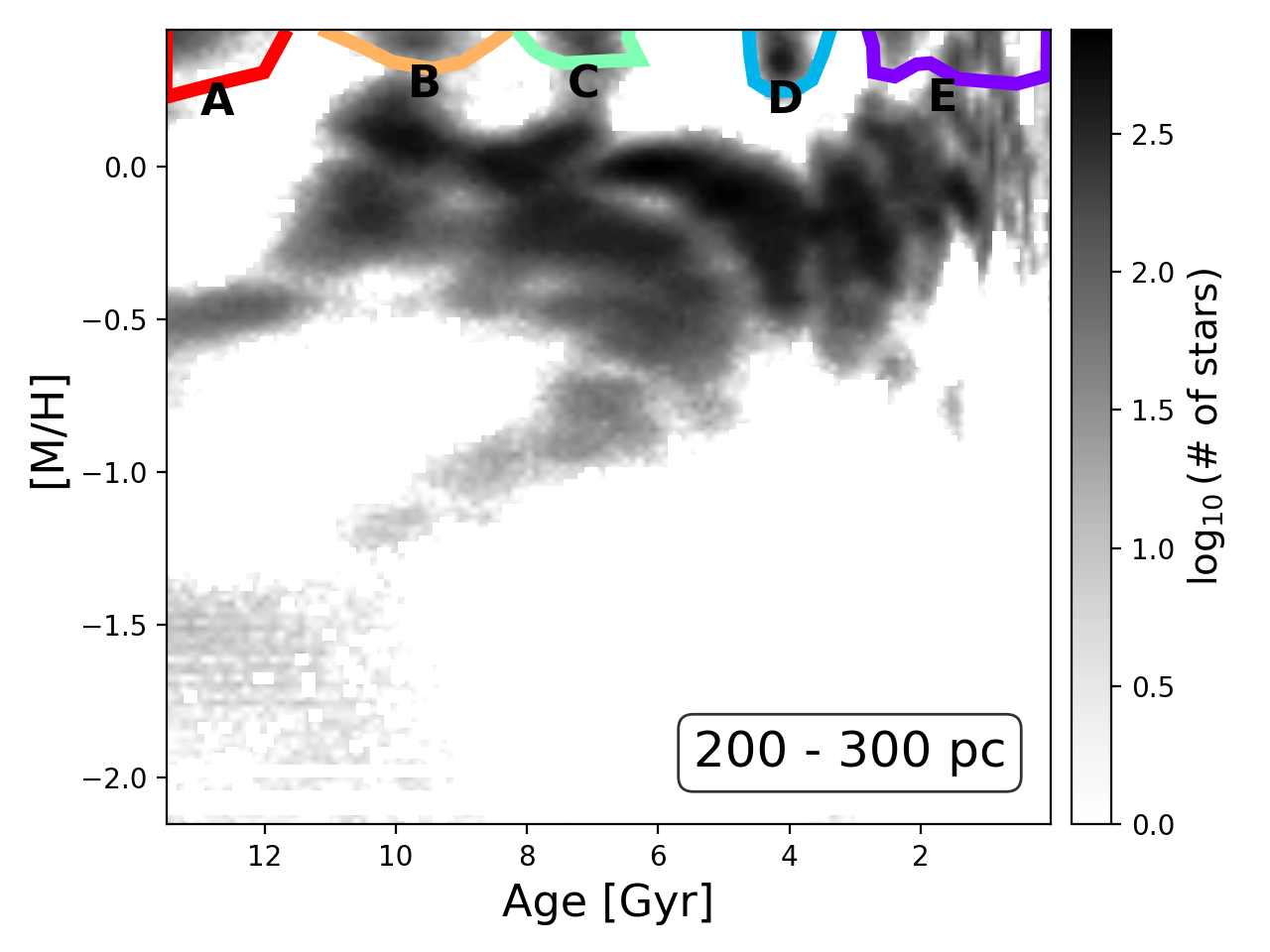} ~
\includegraphics[width=0.32\textwidth]{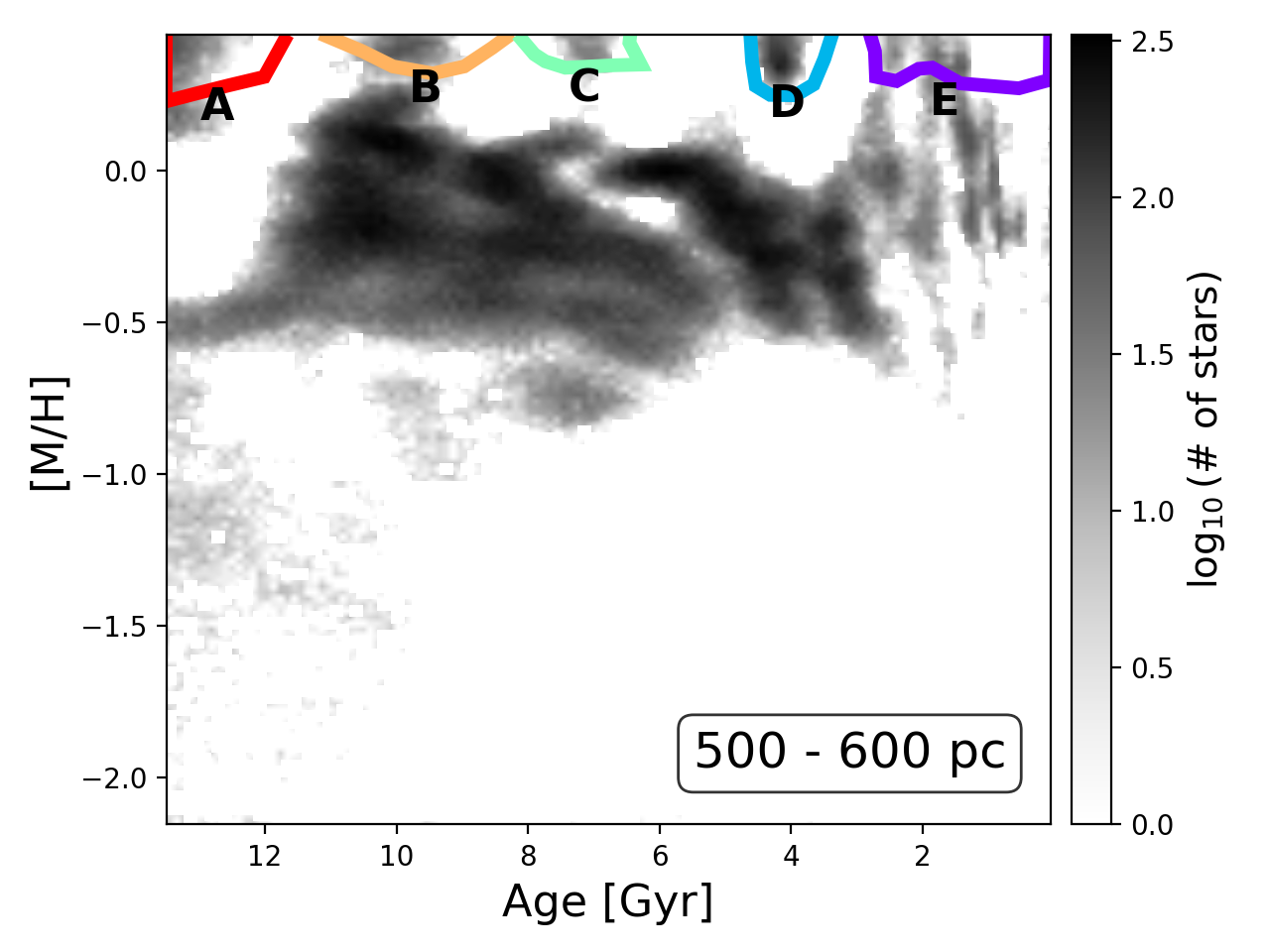} \\
\caption{Stellar density distribution in the age-metallicity plane for volumes 1, 5, and 8 (i.e. from the plane to 50 pc, 0.2 to 0.3 kpc, and 0.5 to 0.6 kpc, respectively above and below the plane) as representative examples of the reported super metal-rich populations (for these we use the \citealt[][]{2022A&A...661A.147L} dust map). Coloured polygons delimit the areas in the age-metallicity plane used to quantify the {\it z}-profiles in Fig.~\ref{fig:prof}. Note that a logarithmic scale has been used to represent the number of stars, in order to enhance these relatively low intensity features.}
\label{fig:SFH_blobs}
\end{figure*}

Fig.~\ref{fig:SFH_blobs} displays the derived distribution of stars in the age-metallicity plane for volumes 1, 5, and 8 using the \citet[][]{2022A&A...661A.147L} dust map. The stellar populations present in these age-metallicity distributions show a smooth variation from volume to volume. Despite this, we find a notable presence of super metal-rich stars in the three exemplary volumes ([M/H]$\sim$0.3-0.45; 9.2, 6.4, and 2.8$\%$ for volumes 1, 5, and 8, respectively). Interestingly, rather than occupying the full age range, they pile up at particular stellar ages, namely $\sim$ 13.5, 10, 7, 4, 2, and younger than 1 Gyr \citep[as already hinted in][see areas A to E highlighted by the coloured contours in this figure]{Gallart2024}. Despite showing some spread in age and metallicity, partly a consequence of an age resolution effect, these results are compatible with the presence of narrow events of metal-rich star formation in the MW \citep[see discussion in][]{Gallart2024}. Also, we should highlight that the relative importance of these stellar populations slightly changes from volume to volume. Interestingly, starting from volume 8 (i.e. 0.5 to 0.6 kpc), these metal-rich populations progressively disappear (or the associated number of stars becomes so low that our method cannot detect them) towards higher $z$. First, population C disappears (volume 10, i.e. 0.7 to 0.8 kpc). Then, population B (volume 13, 1.0 to 1.2 kpc), followed by population D (volume 14, 1.2 to 1.6 kpc). By volume 15, 1.6 kpc and above, none of the populations are detected. Despite showing some spread in age and metallicity, partly a consequence of an age resolution effect, these enhancements are compatible with the presence of narrow events of metal-rich star formation in the MW \citep[see discussion in][]{Gallart2024}. In this work we will focus on such metal-rich stars, leaving the rest of the results (all ages and metallicities) for a separate, in-depth paper dissecting the MW disc. Interestingly, as we will hypothesise in Sect.~\ref{sec:scenario}, the ages of most of the stellar over-densities may coincide with the timing of several MW accretion events. As can be seen from that figure, the metallicity distribution for all these bursts has a sharp limit towards higher metallicities, mainly a consequence of the upper limit of the metallicity grid available in the models ([M/H]=0.45). Thus, we cannot rule out the possibility of having stars of a slightly higher metallicity than that. But, comparisons with spectroscopic metallicities \citep[see][]{alvar2025} suggest that the number of stars with a metallicity higher than our models' upper limit should be very small.

\begin{figure}
\centering
\includegraphics[width=0.45\textwidth]{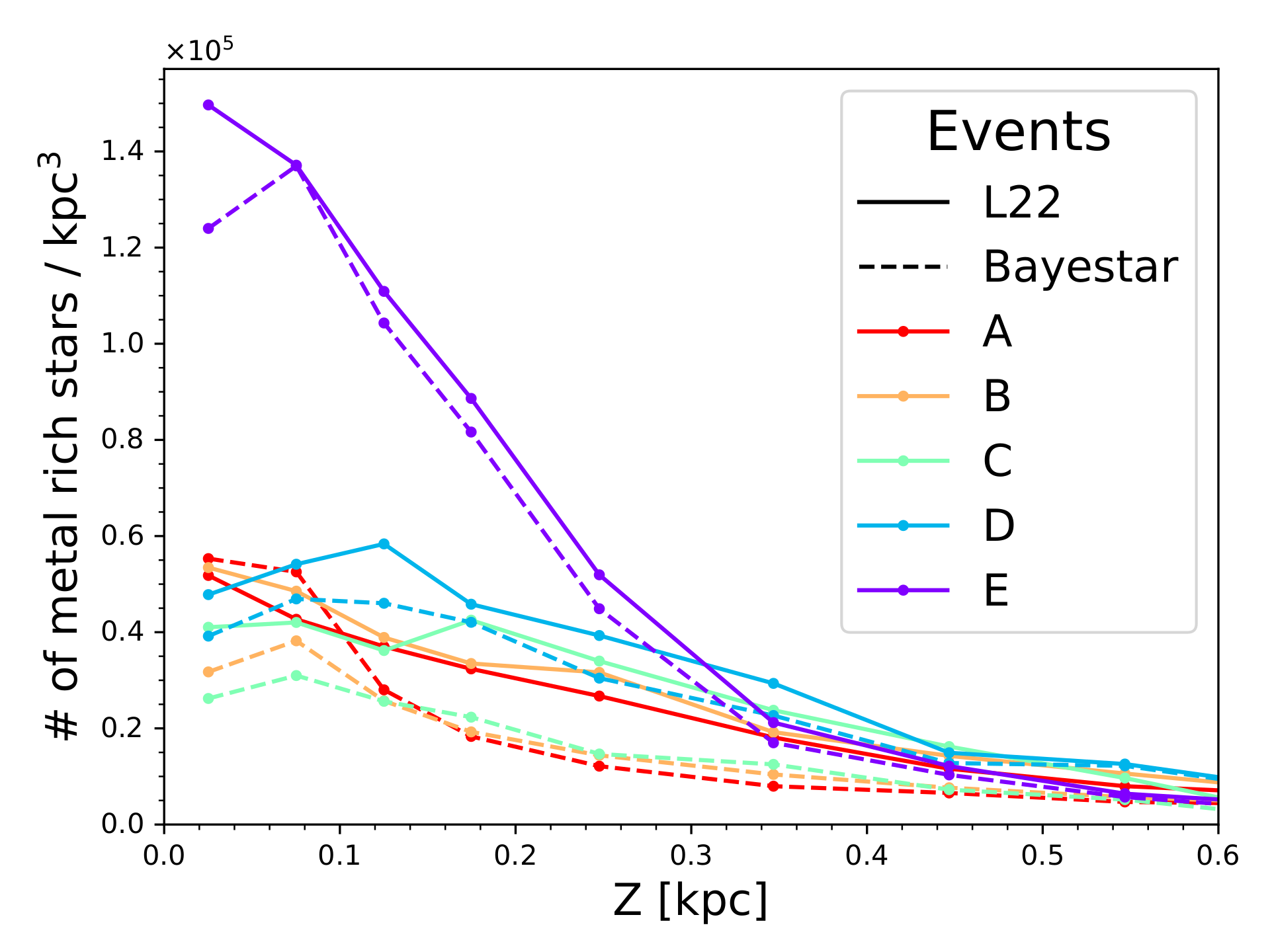} \\
\caption{{\it z}-profile of number density of stars for the five different events of super metal-rich star formation. Such events are defined using the polygons depicted in Fig.~\ref{fig:SFH_blobs}. We show the profiles using two different extinction maps, Bayestar map \citep[][dashed lines]{Green2019}, and \citet[][]{2022A&A...661A.147L} map (solid lines, L22). Given incompleteness affecting the observed samples together with quality cuts, absolute values for this number density should be taken with caution. A normalisation has been applied to the Bayestar densities to account for the missing quadrant in the Bayestar coverage \citep[see][]{Green2019}. Colours follow Fig.~\ref{fig:SFH_blobs}.}
\label{fig:prof}
\end{figure}

Figure~\ref{fig:prof} characterises how the volume density of super metal-rich stars changes as a function of $|z|$ (height above and below the Galactic plane) for each of the identified events (different colours for different events and different lines for different dust maps). From that figure we can conclude that:

\begin{itemize}
    \item We find compatible results using  different dust maps, especially in the case of the youngest events. Note the discrepancy for population E in the inner most region, probably linked with issues with the bayestar dust map.
    \item All events present a declining $z$ profile, although each event declines differently. For instance, event E is the most prominent event until $\sim$~0.3 kpc, where events like C or D become more important.
    \item Beyond 600 pc above and below the plane, the signature of these events nearly disappears (possibly due to the low number of stars and our methodology not being able to detect them).
\end{itemize}  

Given the effects that the quality cuts described in Sect.~\ref{sec:data} have on the completeness, especially in the inner parts, we should state that all these points should be considered as lower limits and that the real profiles should be steeper.

The presence of a large amount of super metal rich stars in the solar neighbourhood is striking by itself. However, its peculiar distribution in stellar age is particularly noteworthy, hinting at the presence of enhanced events of star formation. The smooth variation with height shown in Fig.~\ref{fig:prof} and the agreement between different dust maps rule out the possibility that the finding of these super metal-rich stars is simply the consequence of a deficient reddening correction\footnote{To further check this possibility, we also investigated a different cylinder with a lower radius extent. Totally compatible results are found for all events (including the oldest one) fully ruling out that reddening might artificially create these features.}, although a unlikely missmatch between models and observations might be another option.

\section{Discussion}
\label{sec:discussion}

In this work we find clear evidence of the existence of a super metal-rich stellar population in the solar neighbourhood displaying a peculiar (bursty) age distribution. As we will see in this section, we speculate with the possibility that these metal-rich stars were born in the inner regions of the MW and migrated outward, probably on chaotic orbits due to the influence of the bar. This observational finding has the potential to provide crucial insights on the formation and evolution of our Galaxy's inner parts. But, are these results consistent with the current knowledge? In this section we compare our results with the Auriga Superstars set of simulations as well as put together our observational results (Sect.~\ref{sec:results}) and previous works in the field to propose a feasible scenario on how our Galaxy's inner parts built up.

\subsection{Comparison to a cosmological simulation}
\label{sec:Auriga}

In order to provide some physical insight about the possible origin of the super metal-rich stars, we inspect the Auriga Superstars suite of cosmological simulations. These simulations are an improvement on the Auriga suite of 40 gravo-magnetohydrodynamic cosmological zoom-in simulations of the formation of Milky Way-mass halos \citep{GGM17,Grand2024}, run with the moving-mesh code {\sc arepo}. In this simulation, instead of forming just a single star particle per star-forming gas cell (as usual), superstars forms 64, and metals are injected only in the gas cell in which the star particle is located. For more information on the technicalities of these simulations and how they improve the former Auriga set, we encourage the reader to check \citet[][]{PGG17, GGM17, 2021A&A...650L..16F, Grand2023}; Fragkoudi et al. in prep.; and Pakmor et al. in prep. 

\begin{figure*}
\centering
\includegraphics[width=0.47\textwidth]{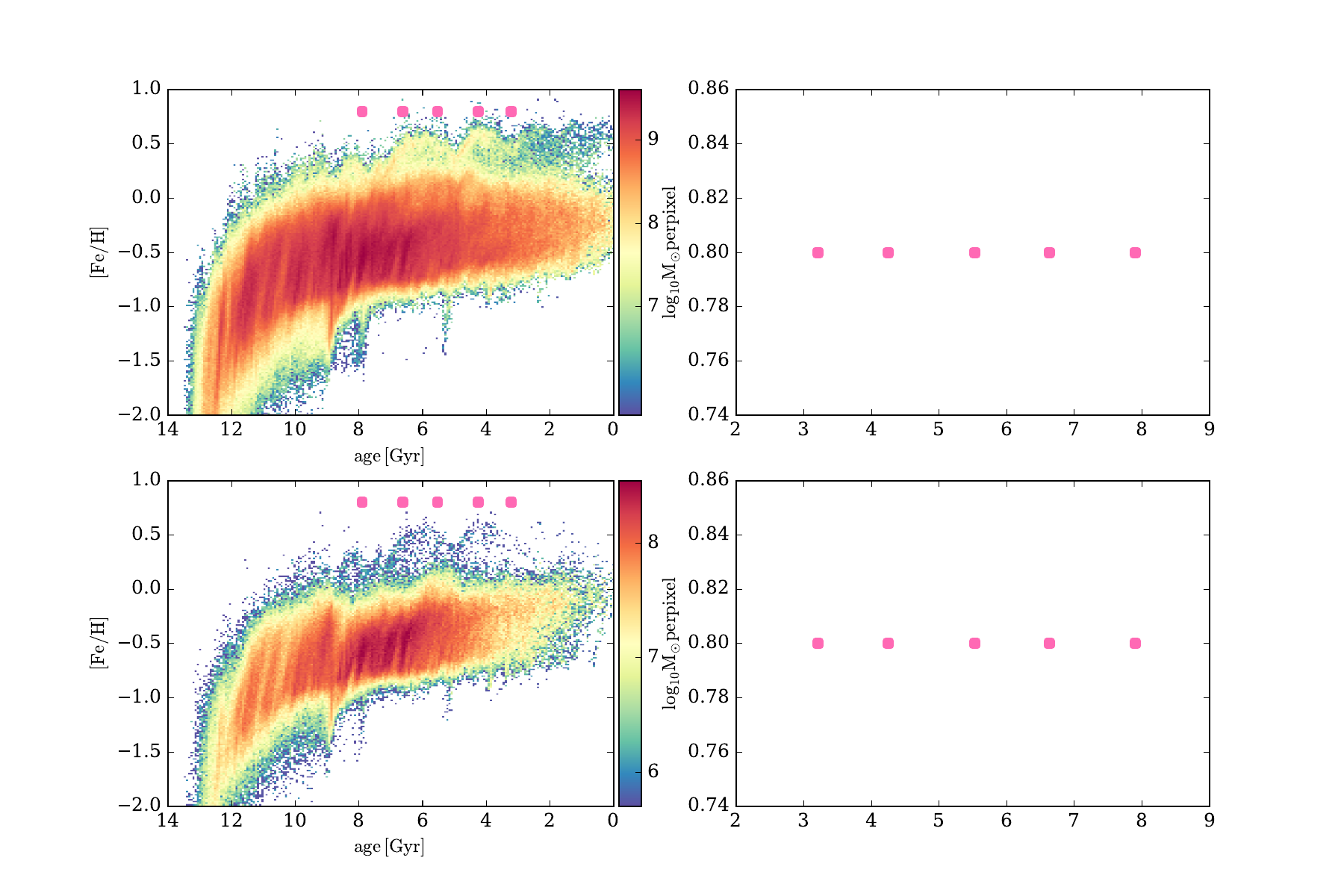} 
\includegraphics[width=0.47\textwidth,]{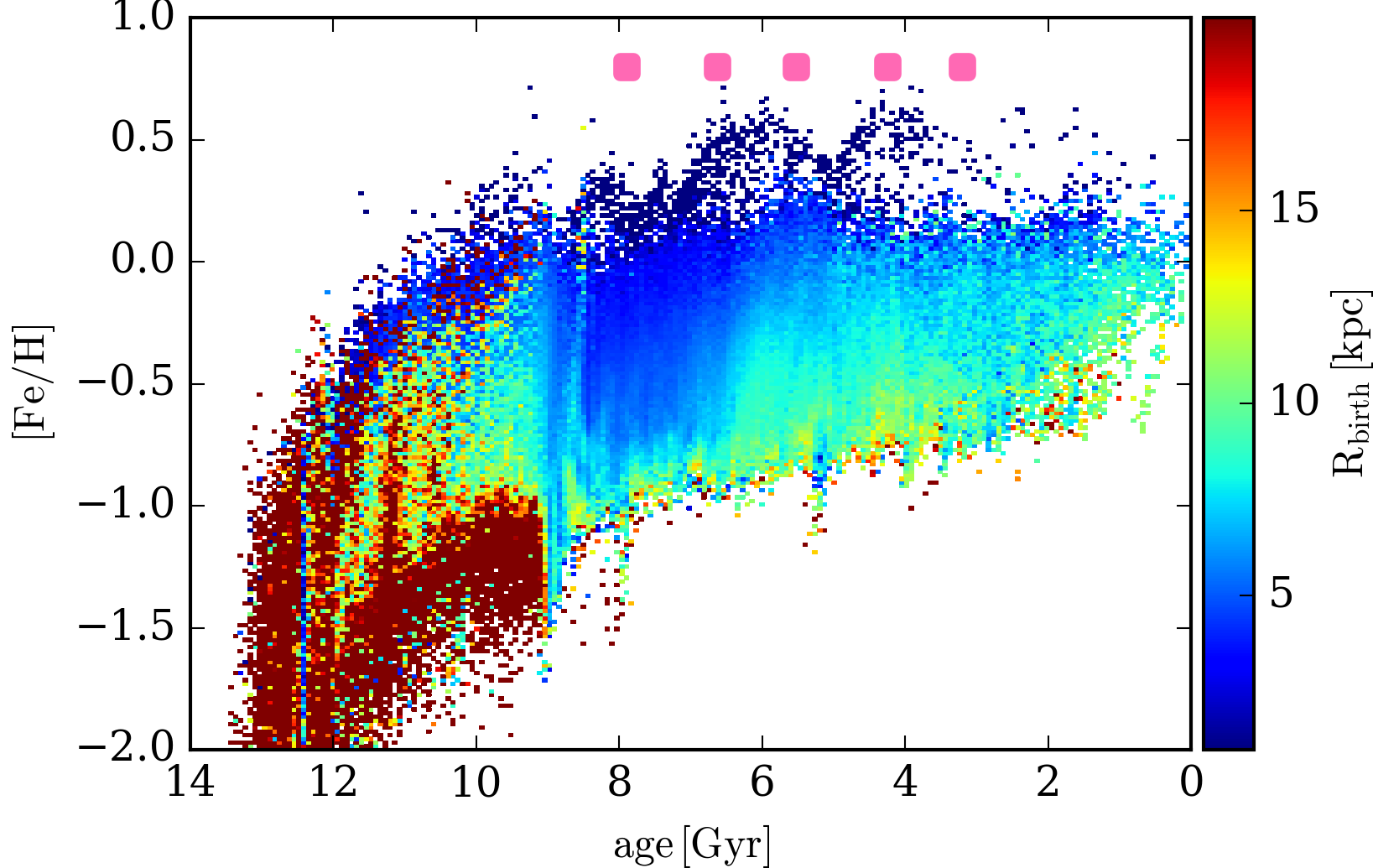} \\
\caption{Stellar age-metallicity distribution of stellar particles for a solar neighbourhood-like selection of stars from AuS18. The distribution of stars are colour-coded according to number density (left) and birth radius (right). Pericentric passages of subhalo 6281 are shown as pink squares.}
\label{fig:fehvsage_subset}
\end{figure*}

Upon inspection of the Auriga Superstars simulations, we find that similar populations of super metal rich stars, with discontinuous age distribution, and somewhat disconnected from the main age-metallicity distribution, can be found in solar neighbourhood-like regions of barred galaxies. As an example, Fig.~\ref{fig:fehvsage_subset} shows the distribution of stars in the age-metallicity\footnote{We subtract 0.4 dex from the iron abundance of the star particles, consistent with previous studies of the Auriga simulations \citep[e.g.][]{Grand2020_dualOriginThickDiskGES}.} plane from a ``solar neighbourhood'' of Auriga Superstars 18 (AuS18), a spiral galaxy hosting a strong bar and a boxy-peanut bulge that can be considered a MW analogue \citep[see][]{Fragkoudi2020bulge} which also includes a Gaia-Enceladus-Sausage-like merging event \citep[e.g.][]{Fattahi19, Merrow2023_barGES, 2024MNRAS.535.2873Z}. In this figure we colour-code the distribution of stars based on stellar density (left) and birth radius of the stars (right). We can find a population of super metal-rich ([Fe/H] above 0.1) stars delineating a discontinuous sequence of ``ups and downs'' in this plane separated from the bulk of the population. Also, based on information on their birth radius (right-hand panel), we can see that all these metal-rich stars (currently located in the solar neighbourhood) were born in the inner parts of AuS18 (inner kpc, mainly the bar, and the boxy-peanut shaped bulge) and migrated outwards. It is interesting to highlight here that, these bursts seem to be related to the pericentric passages of subhalo 6281 (pink squares), one of the most prominent mergers of AuS18\footnote{Some relevant information on this particular halo: log(M$_{\rm Total}$)~=~10.062; log(M$_{\rm \star}$)~=~8.3; infall time: 8 Gyr ago.}. The onset of these ``ups and downs'' coincides with the first pericentric passage of subhalo 6281 ($\sim$~8~Gyr ago), and disappear right before the last approach. The physics behind this phenomenon will be discussed in a future, in-depth paper. The fact that, from all AuS simulated systems, these features only appear in galaxies hosting a bar, and especially the fact that stronger bars (as AuS18) display the clearest signs of them, seem to suggest that the bar plays an important role in shaping what we see (inducing radial migration, funnelling material to the centre enabling metal-righ star formation, etc.). 

The exact mechanisms shaping this population and forcing it to migrate to the solar neighbourhood are unknown and will be subject of study in a separate paper. However, we should advance here (as we will discuss later) that the escape of stars in chaotic orbits from the inner Galaxy to larger radii is a promising possibility to explain our findings \citep[][]{2006A&A...453...39R, 2016MNRAS.457.2583J}. For the purpose of the current work, it is relevant to mention that, from a theoretical point of view, it is expected to find metal-rich stars from the inner Galaxy as far out as the solar neighbourhood, especially in galaxies hosting a bar.

\subsection{Metal-rich stars in the Milky Way and their existence in the solar vicinity}
\label{sec:more_metalrich}

Stars of high metallicities such as the ones discussed in this work have been known to exist in the MW for a long time \citep[e.g.][]{1972ade..coll...55G, Kordopatis2015RichAreDifferent, MiglioChiappini2021_Kepler}. In particular, the MW bulge is a complex superposition of stellar populations of very different metallicities \citep[including super metal-rich stars, see e.g.][]{Zoccali2008, ness_freeman2016, Zoccali2017_GIBSIII,  2022A&A...666A..72N, queiroz21}. However, these metal-rich stars are not only present in the innermost regions of the MW, but also, in a lower number, in the solar neighbourhood and along the Galactic plane \citep[e.g.][]{2011A&A...535A..42T}. Whether these stars were born in the solar vicinity or not is not entirely known. Most evidence suggests that these stars are not formed locally, but have migrated from the inner parts \citep[][]{Halle2015churblur, hayden2015, haywood2019, 2020A&A...638A.144K, 2023A&A...669A..96D, Nepal2024_YoungBar}. To fully unveil their origin, a complete description of their properties, including dynamics, chemistry, and ages (to compare with our results), is needed. 

Deriving stellar ages is an especially difficult task \citep[][]{soderblom2010}. The most common method to obtain age information in Galactic archaeology is through Bayesian isochrone-fitting methods (e.g. {\tt STARHORSE}, \citealt{Queiroz2023_starhorse}) using photometric, astrometric and spectroscopic data simultaneously to derive ages of individual stars. The addition of asteroseismology data constraining stellar masses can lead to an improvement in the age estimates \citep[][]{1986ApJ...306L..37U, Miglio2017_Plato, 2019A&A...622A.130B}. \citet[][]{Nepal2024_YoungBar}, using {\tt STARHORSE} stellar ages, identified metal-rich stars as young as $\sim$~3 Gyr (bulk 6-11 Gyr)  in the solar neighbourhood and used them to argue for a recent formation of the MW bar \citep[see also][for evidence of young metal-rich stars in the solar neighbourhood]{haywood2013, MiglioChiappini2021_Kepler}. Thus, the study presented here agrees with previous works on the existence of metal-rich stars with a wide age range in the solar neighbourhood. However, given that the consensus seems to indicate that these metal-rich stars migrated from the central parts, the real question we are tackling is actualy related to the ages of the stars in the inner MW.

\citet[][]{zoccali2003}, combining optical and near-IR stellar photometry, simultaneously quantified the metallicity distribution, age, and luminosity function of the Galactic bulge stellar population with no trace of any population younger than $\sim$10~Gyr. But this is just the beginning of a myriad of works supporting for an eminently old bulge based mainly on Hubble Space Telescope deep CMDs. \citet[][]{2008ApJ...684.1110C} analysed the tightness of the oldest-main sequence turn-off stars in the SWEEPS field \citep[][]{2006AAS...20916221S} to conclude that the stellar populations in the bulge are $\sim$ 11 Gyr old, and no younger than 8 Gyr old \citep[although see][where the authors conclude that a wide range of stellar ages can also give a tight turn-off]{haywood2016bulge}. In the same line, \citet[][]{renzini2018}, analysing both CMDs and luminosity functions, set an upper limit on the amount of young stars in the bulge, featuring an insignificant population of stars younger than  $\sim$5~Gyr. However, the careful spectroscopic analysis by \citet[][]{bensby2013, bensby2017}, more successful at breaking the age-metallicity degeneracy, suggested the presence of stars of almost all ages in the bulge. In particular, the authors found signs of a bursty behaviour, with episodes of enhanced star formation 3, 6, 8, and 11~Gyr ago. In addition, evidence exists suggesting that the nuclear stellar disc may have experienced a sudden burst of star formation in the last Gyr \citep[][]{nogueras-lara2020burst}. From the theoretical side, numerical simulations of barred, disc galaxies also find that the star formation in the bars can proceed in a bursty way, due to feedback or triggered by interactions and satellite flybys \citep[e.g.][]{seo2019, bi2024}. Moreover, several works have claimed that the MW bulge stellar population is in fact a mixture of thin and thick disc stars highly influenced and trapped by the dynamics of the bar \citep[e.g.][]{2016PASA...33...27D, 2017A&A...606A..47F, 2017MNRAS.469.1587D}. For a recent, comprehensive review on what is known about the MW bulge see \citet[][]{2024arXiv241201607Z}. All these works, theoretical and observational, support the existence of a patchy stellar age distribution for the bulge stars, in agreement with our findings if we assume that the super metal-rich stars in the solar neighbourhood formed in the inner MW and migrated outwards. 

\subsection{The Gaia GSP-Spec perspective}
\label{sec:recio}

\begin{figure*}
\centering
\includegraphics[width=0.95\textwidth]{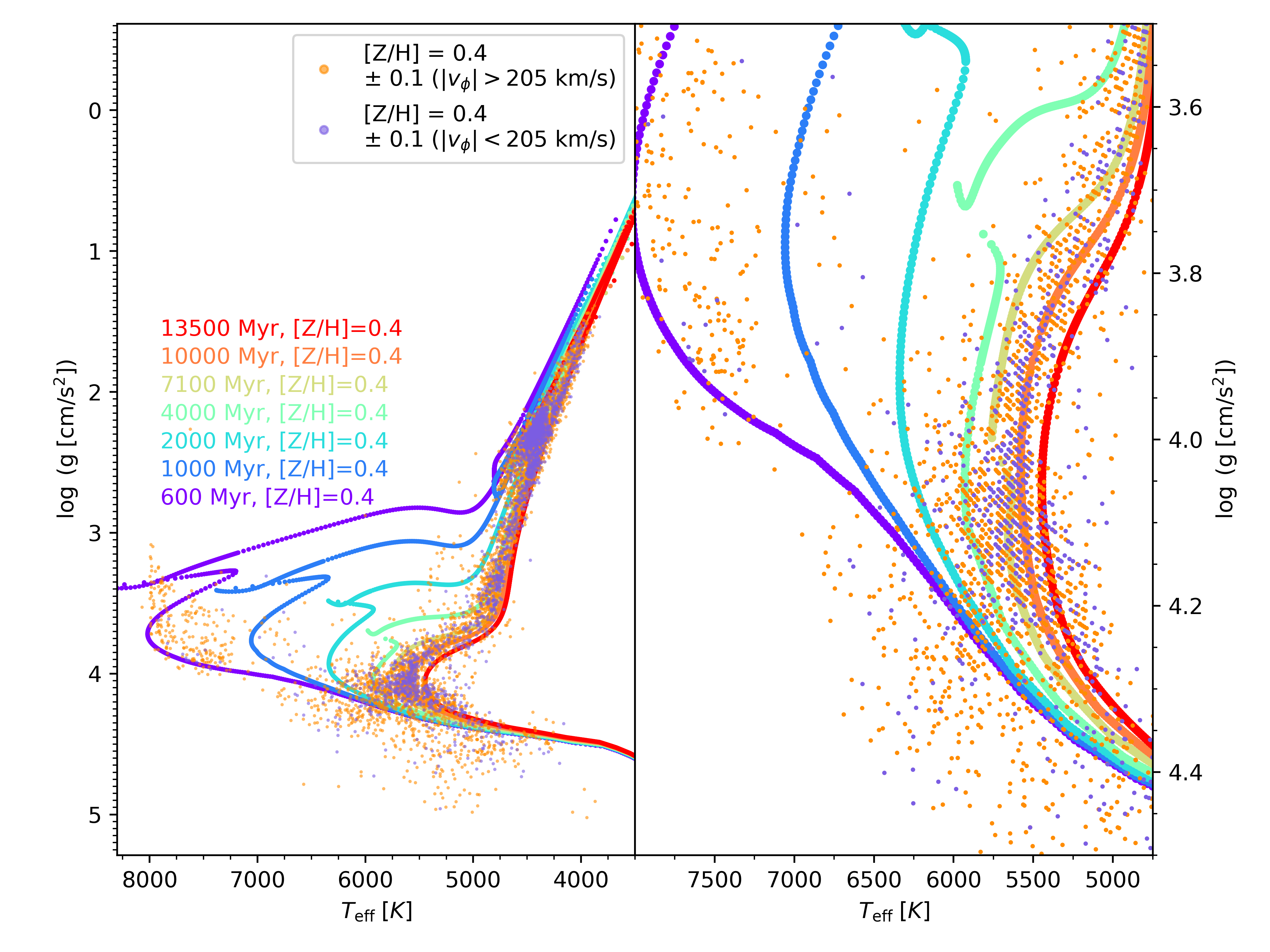} \\
\caption{Kiel diagram of the disc super metal-rich population with [M/H]=0.4$\pm$0.1 dex. We divide the sample in two based on their value of v$_\phi$: slow (purple) and fast (orange) stars (see Fig.~\ref{fig:vphi}). Solar-scaled BaSTI isochrones of 13.5, 10, 7, 4, 2, 1, and 0.6 Gyr (red, orange, pistacho, green, cyan, blue, and purple, respectively) are overlaid on the data. {\it Left:} Whole diagram. {\it Right:} Focusing only on the turn-off region.}
\label{fig:metalrich_kiel}
\end{figure*}

\citet[][]{alejandra2024}, using a selection of high-quality stellar chemophysical parameters from the Gaia DR3 GSP-Spec catalogue  \citep[][]{2023A&A...674A..38G}, also reported the presence of super metal-rich stars in a wide age range in the solar neighbourhood. In addition, they clearly observe two populations of giant stars in the Kiel diagram (log(g) vs. T$_{\rm eff}$; see their figures 8 and 11), that they associate with the thin vs. thick disc bimodality \citep[a complete discussion on MW thin vs. thick discs can be found in][]{kawata_chiappini2016}. After comparing with BaSTI isochrones, they demonstrate the need of an age gap in order to explain the separation of both evolutionary sequences as well as the distribution of stars near the main sequence turnoff. To put our results within the context of their analysis, and to see if their reported age gap is consistent with the distinct episodes of star formation inferred from our derived age-metallicity distribution, Fig.~\ref{fig:metalrich_kiel} reproduces their figure 6 and expands on it. 

Figure~\ref{fig:metalrich_kiel} shows a Kiel diagram of the sample of stars within our volume analysed in \citet[][]{alejandra2024} with metallicities\footnote{We note that our results are given in terms of global metallicities ([M/H]), so in principle the iron abundance estimates by \citet[][]{alejandra2024} should be transformed to [M/H] by accounting for the corresponding [$\alpha$/Fe] and a rescaling law \citep[see e.g.][]{Salaris1993_feh_to_mh}. However, since the selected stellar sample corresponds to metal-rich stars with a very low - if any - $\alpha$-element enhancement, we decided to not apply any transformation for the purpose of the present comparison, so assuming that [Fe/H] $\sim$ [M/H] $\sim$ [Z/H] (global metallicity, as appropriate for solar-scaled stars).} in the range [Fe/H] = 0.3-0.5. We overplot some BaSTI isochrones coinciding with our main detected metal-rich populations (13.5, 10, 7, 4, 2, 1, and 0.6 Gyr). By matching the position of the stars with these isochrones, we can confirm the presence of metal-rich stars of ages even younger than $\sim$ 2 Gyr old (in agreement with our stellar age distribution). The right-hand panel in Fig.~\ref{fig:metalrich_kiel} (zooming into the main sequence turn off region) highlights how well all the old and intermediate age isochrones (13.5, 10.0, 7, and 4 Gyr) match the main sequence turnoff and subgiant branch. Thus, this plot confirms our findings on one of the highest-quality sets of stellar astrophysical parameters to date. Not only are there metal-rich stars in the solar neighborhood, but they also exhibit a wide range of ages and a noticeable discretization (not stars of all ages) that is consistent with our episodic stellar age distribution. Nevertheless, we need to add that, although this comparison validates our results, a study such as the one presented in this paper is necessary to achieve the age resolution required \citep[][]{Gallart2024} to fully characterise the age distribution of metal-rich stars in the solar neighbourhood. It also enables us to quantify the relative importance of each population: note that this particular spectroscopic sample has strong selection effects derived from the requirement of highly accurate measurements of the stellar parameters.

\begin{figure*}
\centering
\includegraphics[width=0.31\textwidth]{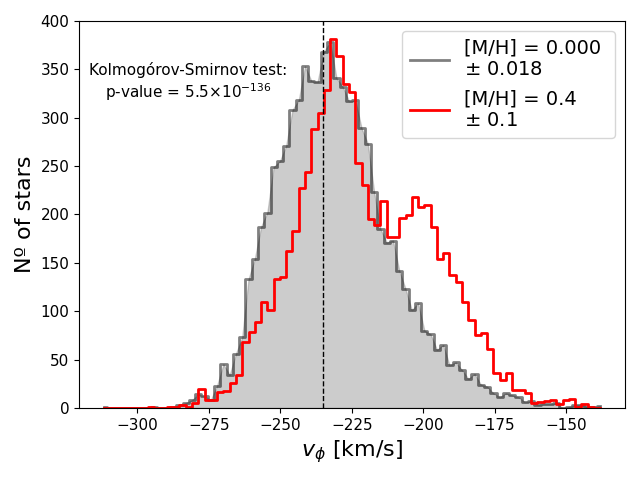} ~
\includegraphics[width=0.31\textwidth]{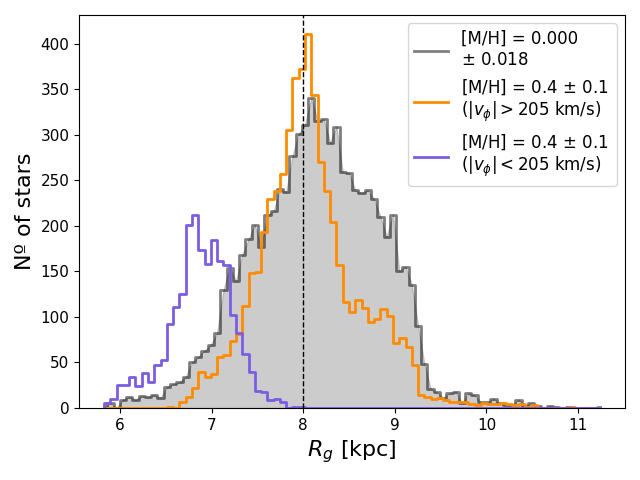} ~
\includegraphics[width=0.31\textwidth]{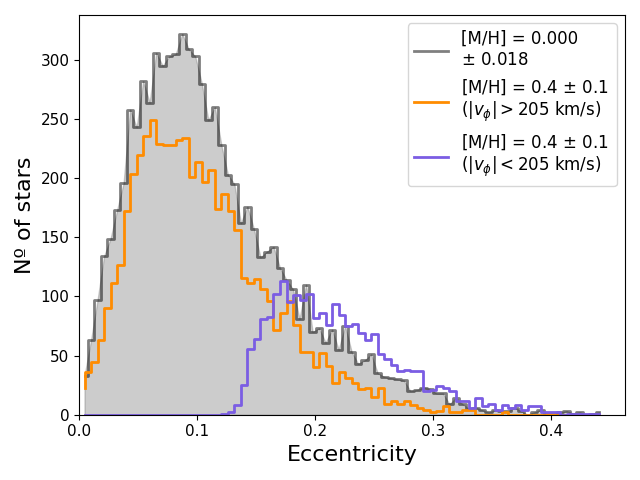} \\
\caption{Orbital properties of the metal-rich stars detected in the high-quality Gaia DR3 GSP-Spec sub-catalogue from \citet[][]{alejandra2024}, compared with those of a subset of solar-metallicity stars. {\it Left:} Distribution of v$_\phi$ velocities for metal-rich (red, empty histogram) and solar-metallicity (grey histogram) stars. From the shape of the metal-rich stars histogram we divide the sample in slow (|v$_\phi$| below 205 km/s) and fast (|v$_\phi$| above 205 km/s) stars. {\it Middle:} Distribution of guiding radius \citep[from][]{alejandra2024}. {\it Right:} Distribution of eccentricity \citep[from][]{alejandra2024}. For these last panels, we divide the sample in solar metallicity (grey), slow, metal-rich stars (purple) and fast, metal-rich stars (orange).}
\label{fig:vphi}
\end{figure*}

In order to further assess the origin of the metal-rich stars currently observed in the solar vicinity, Fig.~\ref{fig:vphi} compares the orbital properties (v$_\phi$, guiding radius, and eccentricity)\footnote{All dynamical parameters have been taken from \citet[][]{alejandra2024}, with the exception of guiding radius that has been computed as the average of the apocentric and pericentric radii, i.e. (R$_{\rm peri}$+R$_{\rm apo}$)/2. For more details see \citet[][]{palicio2023}} for metal-rich stars and a similar number of solar-metallicity stars\footnote{We adjust a range of metallicities around the solar value to have about the same number of stars in the solar regime as in the high metallicity range defined above. To be precise, we consider a star to be a solar-metallicity star if its metallicity is in the range 0.0000$\pm$0.0175 ([Fe/H]).} from \citet[][]{alejandra2024}. From the figure (left panel), we can identify two distinct families of metal-rich stars. On the one hand, there are some stars that present lower values of |v$_\phi$|, lower guiding radii, and larger eccentricity (purple colours in Fig.~\ref{fig:metalrich_kiel}, and Fig.~\ref{fig:vphi}) with respect to the overall solar-metallicity population. We call these metal-rich stars slow stars. On the other hand, we also have stars with very similar orbital properties as the solar-metallicity stars in the solar neighbourhood (in orange in Fig. ~\ref{fig:metalrich_kiel} and in the middle and right-hand panels of Fig. \ref{fig:vphi}). We call these metal-rich stars fast stars. As we will see, these two behaviours are consistent with different radial migration mechanisms.

\citet[][]{schonrich2009} proposed the existence of two different drivers for radial migration, namely churning (change in angular momentum manifested as a change in guiding radius, also known as diffusion) and blurring or radial heating (a change in the epyciclic amplitude leading to epyciclic excursions). Thus, fast stars, sharing orbital properties with the bulk of the population near the Sun, might be examples of stars that migrated from the inner regions via churning, and thus, share orbital properties with stars born in the solar vicinity. On the other hand, slow stars might be in their apocentres (explaining the low value of |v$_\phi$|), just experiencing an epyciclic immersion to the outer parts of their orbits (high eccentricity and lower guiding radius). All this is consistent with super metal-rich stars being present at the solar radius via both migration mechanisms, blurring and churning. 
Interestingly, while fast stars are around 2.5 times more abundant than slow stars overall, in the temperature range from 7000 to 8000 K (see Fig.~\ref{fig:metalrich_kiel}, dominated by stars younger than 1 Gyr), the ratio is $\sim$ 11, highlighting a near absence of young, slow stars migrated via blurring. This can be seen as young metal-rich stars being present in the solar neighbourhood preferentially via churning, which could be interpreted as churning being more efficient than blurring in terms of radial distance migrated per Gyr (i.e. young stars migrating due to blurring did not have time to get to the solar radius). This would be in agreement with \citet[][]{2020ApJ...896...15F}, who concluded, using APOGEE data, that radial migration in the MW disc is dominated by diffusion in angular momentum, i.e. churning. However, a migration speed of the order of a solar radius per Gyr is probably unfeasible \citep[see e.g.][]{2022MNRAS.511.5639L}, and thus, we have to leave the door open to the possibility that very young, metal-rich stars are really born in situ.

After all the discussion so far, it is reasonable to think that the discontinuous age distribution that we find for metal-rich stars in the solar neighbourhood could be mirroring the age distribution of stars in the innermost regions of our Galaxy. It is now the moment to put all these pieces together to form a comprehensive, evolutionary picture.
 
\subsection{Reconstructing the past of the MW inner regions}
\label{sec:scenario}

The discretization in stellar age displayed by metal-rich stars unveiled in this work reflects on possible bursts of star formation $\sim$ 13.5, 10, 7, 4 and less than 2 Gyr ago. Among the many mechanisms that could trigger the formation of new stars, merger events and interaction with satellites stand out \citep[][]{mihos1994, hernquist1995, dicintio21, Renaud2021_VINTERGATAN_II, Renaud2021_VINTERGATAN_III, orkney22}. In particular, in the case of barred galaxies, as the bar could act as a conveyor belt driving gas through the bar to the inner region \citep[e.g.][]{2016MNRAS.462L..41F, perez2017, seo2019}, these newly born stars could be formed from gas compressed in the centre by tidal forces from the satellite passages. Putting together this and our current knowledge on the MW accretion history, a plausible, yet speculative, scenario on the past of the MW inner regions appears.

The MW would initially form stars rapidly, resulting in the formation of the oldest stars in our Galaxy, with ages $\sim$13.5~Gyr. The intensity of the star formation at that time would result in the formation of very metal-rich stars early in the history of the Universe \citep[e.g][]{2011A&A...535A..42T}. After this, many small accretion events should have taken place \citep[e.g.][]{2020ARA&A..58..205H}, with Gaia-Sausage-Enceladus \citep[merging around 10-9 Gyr ago][]{Gallart2019Gaia, diMatteo2019noInSitu, Montalban2021GES} being the latest major accretion event experienced by the MW \citep[][]{belokurov2018_sausage, helmi2018, Ciuca2024}, and possibly triggering the formation of the bar \citep[][]{Merrow2023_barGES}. It was not until $\sim$ 7-6 Gyr ago, that the Sagittarius dwarf galaxy \citep[Sgr,][]{ibata94} experienced its first pericentric passage about the MW \citep[][]{law2010, Ruiz-Lara2020Sgr}, recurrently undergoing pericentric passages and affecting the dynamics of the MW from then on \citep[][]{gomez2013, laporte2019, Antoja2020Sgr}. Finally, during the last 2 Gyr or so the MW should have experienced the combined effect of Sgr and the Magellanic system infall \citep[][]{Besla2007FirstInfallMC, laporte2018, patel2020, vasiliev21_Tango4Three}. As the reader may notice, the correlation between the timing of the expected episodes of interaction between the MW and its satellite system and our tentative star formation bursts ($\sim$ 13.5, 10, 7, 4 and less than 2 Gyr ago) is striking. 

\begin{figure}
\centering
\includegraphics[width=0.47\textwidth]{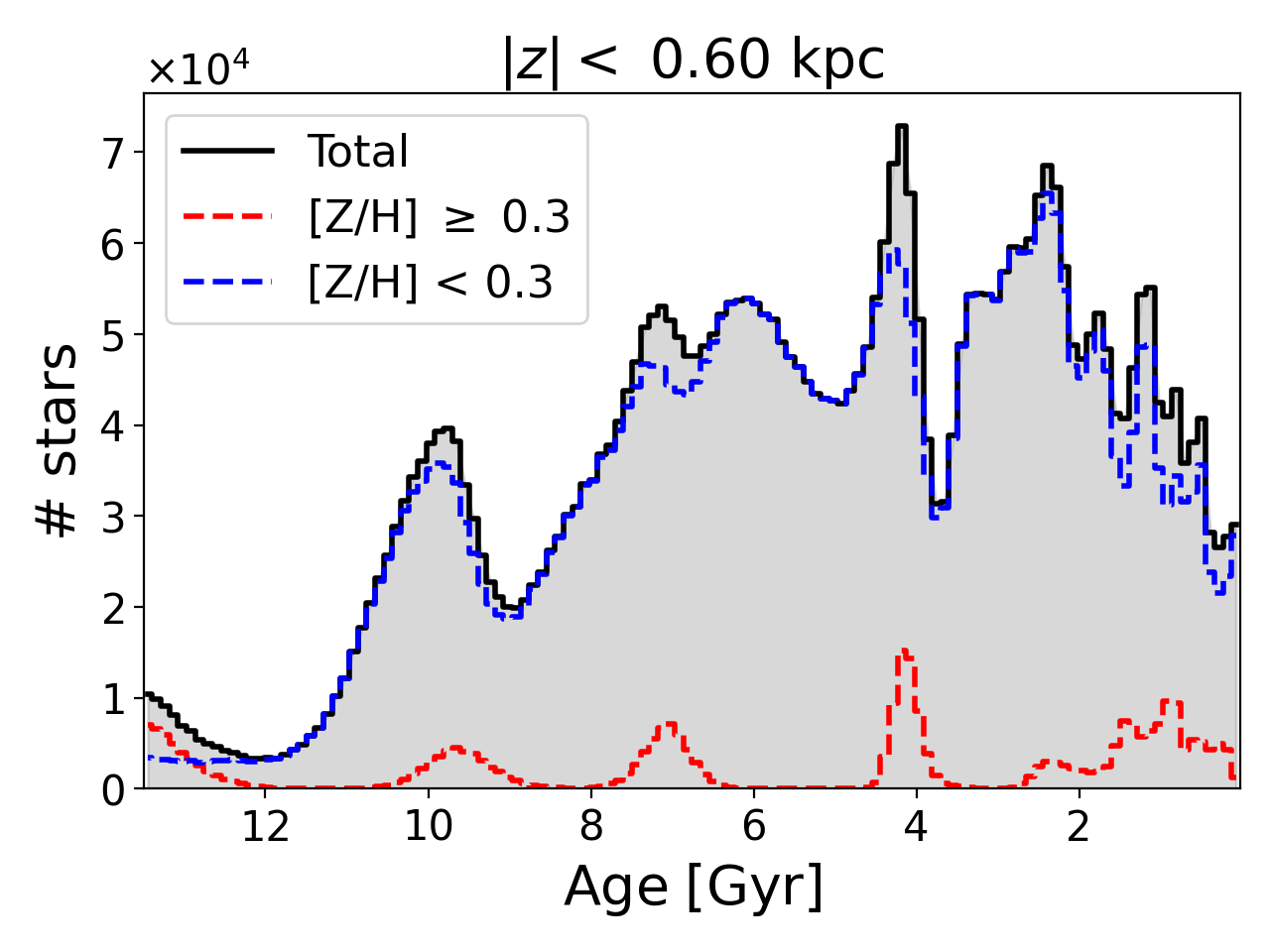} \\
\caption{Integrated stellar age distribution of the stars within 0.6 kpc of the plane of the disc (|z|<0.6 kpc). We divide it into total (including all stars, shaded black area); metal-rich (considering only stars with [M/H] $\geq$ 0.3, red); and rest (stars with [M/H] < 0.3, blue).}
\label{fig:metal_rich_vs_poor}
\end{figure}

Note that these star formation enhancements linked to interactions should not be restricted to the metal-rich stars or concentrated only on the inner Galaxy. They could be global accross the whole MW. From Fig.~\ref{fig:metal_rich_vs_poor}, where we represent the stellar age distribution for stars with [M/H] above and below 0.3 (red and blue, respectively for all volumes within 0.6 kpc from the plane of the disc), it can be noticed that these epochs coincide with the presence of stars and peaks in its numbers at lower metallicities as well. Interestingly, all peaks in the stellar ages of the metal-rich stars seem to be slightly shifted towards younger ages. This might be a consequence of a delay in the star formation from the outer disc to the inner Galaxy. Given that the star formation enhancements are not restricted to the metal-rich end, we can conclude that indeed, we are witnessing global star formation enhancements in the MW rather than central ones.

Therefore, all these external events seem to have induced enhanced star formation in the whole MW, including its central parts \citep[][]{mihos1994, hernquist1995}, that would lead to a fast chemical enrichment, especially in the inner Galaxy. In the case of barred galaxies like our MW (as shown by the comparison with the Auriga Superstars simulations), some mechanisms (probably bar-(spiral)-induced radial redistribution) could move inner stars (of any metallicity) to the solar neighbourhood \citep[e.g.][]{MinchevFamaey2010, iles2024}, with those displaying the highest metallicities standing out due to the lack of in-situ, metal-rich stars in these outer regions. However, there is one difficulty with this scenario. Stars born in the inner parts, in order to get to the solar vicinity, need to pass co-rotation, which, dynamically speaking is not preferred \citep[][]{ceverino2007, Halle2015churblur}. Nevertheless, it has been shown numerically that the bar should affect the kinematics of stars in the solar neighbourhood \citep[][]{brunetti2011, haywood2024}. Also, although only to a limited number, stars in chaotic orbits can cross co-rotation and get to the solar neighbourhood \citep[see Fig.~\ref{fig:fehvsage_subset} but also][]{raboud1998, fux2001}. In particular, the manifold theory of spiral structure \citep[][]{2006A&A...453...39R, 2006MNRAS.373..280V} provides a dynamical mechanisms on how chaotic motions can significantly change the spatial distribution of matter up to the outer regions. Other mechanisms have been claimed to allow inner stars to migrate to the solar neighbourhood such as overlap between bar and spiral arm resonances \citep[e.g.][]{MinchevFamaey2010} or bar slowdown \citep[][]{Chiba2021}. \citet[][]{2022MNRAS.514.5085O}, analysing Auriga simulations, found that strongly barred galaxies show larger stellar migration, though with a slower timescale than diffusion (churning). However, the authors find stars with a net radial migration over time that can reach a maximum $\Delta$R of $\sim$ 5kpc, with rates that can be as high as 25 kpc/Gyr. Thus, as shown in the case of the Auriga Superstars simulations commented in Sect.~\ref{sec:Auriga}, we can reasonably expect to find metal-rich stars from the inner Galaxy ($\sim$ 3 kpc) in the solar neighbourhood, in spite of the possible co-rotation barrier, as soon as 1 Gyr after its formation. 

In this work, we hypothesize that we might be indirectly witnessing how star formation proceeded in the central regions of the MW from the age distribution of metal-rich stars in the solar neighbourhood. These stars are expected to have migrated here via both churning and blurring. However, there are still a couple of aspects to be addressed, namely the absence of metal-rich stars with ages between 6.5 to 4.5 Gyr ago when we do observe a peak of star formation at the solar radius, and the origin of the particularly narrow 4 Gyr old population. The former might suggest a period of decreased or null star formation in the central MW \citep[see][for quiescent periods of star formation in the MW nuclear disc]{nogueras-lara2020burst} as, possibly, the first pericentric passage of Sgr mainly triggered star formation in the mid-to-outer parts of the Galaxy. This interpretation is in line with what was found in \citet[][]{Renaud2021_VINTERGATAN_III, Renaud2021_VINTERGATAN_II} using the Vintergatan simulation \citep[][]{Agertz2021_VINTERGATAN_I}. Further supporting this idea is the presence of a very metal-poor population ([M/H]$\sim$-0.9 dex, not found at any other ages) at an age roughly coinciding with event C. 

Regarding the latter (the 4 Gyr --event D-- enhancement), to our knowledge no merging events are suspected in the MW at this age. However, this population is so clearly detected that it forces us to think that something quite dramatic might have happened 4~Gyr ago in the history of our Galaxy to form such feature. However, the census of MW merging events is far from complete, especially considering the possibility of satellites in orbits near the plane of the Galaxy, whose tidal streams are more difficult to detect \citep[as it is the case of the Icarus stellar stream][]{2021ApJ...907L..16R}. Also, the possible orbit of Sgr after its merger, as far back as 4 Gyr is still uncertain \citep{vasiliev21_Tango4Three}. Interestingly, the 4 Gyr enhancement coincides with what \citet[][]{Nepal2024_YoungBar} suggest to be a period of enhanced star formation due to a high bar activity period, that the authors linked to its formation. However, although our findings do not rule out bar formation $\sim$ 3 to 4 Gyr ago, they also agree with a formation 10 Gyr ago, probably linked to the interaction with Gaia-Enceladus-Sausage \citep[][]{bovy2019, Merrow2023_barGES, sanders2024}. Also, putting this event within the context of the full reconstructed SFH, we see that it is directly followed by a period of low star formation at lower metallicities (in the solar neighbourhood), after which we find another intense period (see Fig.~\ref{fig:SFH_blobs}, labelled as E, see also Fig.~\ref{fig:metal_rich_vs_poor}). This could be related to a second pericentric passage of Sgr or the accretion of a gaseous intergalactic filament associated with the accretion of Sgr \citep[in the line of what is described in][]{Renaud2021_VINTERGATAN_II}. Whatever might have happened to the MW 4 Gyr ago, clearly deserves further investigation beyond the scope of this paper.

\section{Conclusions}
\label{sec:conclusions}

Using CMDft.Gaia, a CMD-fitting technique tailored to Gaia data, we have identified a discrete age distribution for the metal-rich stars present in the solar neighbourhood. By combining these findings with the analysis of Auriga Superstars simulations, ultra-precise stellar chemophysical parameters \citep[][]{RecioBlanco2023_GaiaDR3} and literature works, we piece together a comprehensive, yet speculative scenario to be confirmed by upcoming theoretical works. As these metal-rich stars are unlikely to have formed in the solar vicinity, we use them as tracers of the stellar age distribution present in the inner parts of our Galaxy. The discretization of the stellar age distribution might come as a consequence of global star formation proceeding episodically, driven by external events (early acretion, Gaia-Sausage-Enceladus, Sgr, and Magellanic Clouds). In this scenario, bar-induced radial migration would be responsible for the presence of these metal-rich stars at solar radii although in a low number as passing co-rotation is difficult, but feasible, in dynamical terms.

This work perfectly exemplifies the power of CMD reconstruction techniques to answer and to propose questions about the formation and evolution of our Galaxy. The high-quality and high-resolution age-metallicity distributions that the ChronoGal project is providing and will provide are set to be a milestone in the field.

\begin{acknowledgements}

We are thankful to the anonymous reviewer for improving the quality of the original manuscript. The authors are grateful to C. Chiappini, D. Kawata, A. Recio-Blanco, and M. Zoccali for useful conversations. We acknowledge financial support by the research projects PID2020-113689GB-I00, PID2020-114414GB-I00, and PID2023-150319NB-C21 financed by MCIN/AEI/10.13039/501100011033, the project A-FQM-510-UGR20 financed from FEDER/Junta de Andaluc\'ia-Consejer\'ia de Transformaci\'on Econ\'omica, Industria, Conocimiento y Universidades/Proyecto and by the grants P20-00334 and FQM108, financed by the Junta de Andaluc\'ia (Spain). TRL acknowledges support from Juan de la Cierva fellowship (IJC2020-043742-I) and Ram\'on y Cajal fellowship (RYC2023-043063-I, financed by MCIU/AEI/10.13039/501100011033 and by the FSE+). FvdV is supported by a Royal Society University Research Fellowship (URF\textbackslash R1\textbackslash 191703 and URF\textbackslash R\textbackslash 241005). EFA acknowledges support from HORIZON TMA MSCA Postdoctoral Fellowships Project TEMPOS, number 101066193, call HORIZON-MSCA-2021-PF-01, by the European Research Executive Agency. ABQa acknowledges support from Juan de la Cierva fellowship (PLease add reference). SC acknowledges financial support from PRIN-MIUR-22: CHRONOS: adjusting the clock(s) to unveil the CHRONO-chemo-dynamical Structure of the Galaxy” (PI: S. Cassisi) funded by the European Union - Next Generation EU, Theory grant INAF 2023 (PI: S. Cassisi), and the Large Grant INAF 2023 MOVIE (PI: M. Marconi). RG is supported by an STFC Ernest Rutherford Fellowship (ST/W003643/1). RB is supported by the SNSF through the Ambizione Grant \verb|PZ00P2_223532|. FAG acknowledges support from the ANID BASAL project FB210003, from the ANID FONDECYT Regular grants 1251493, and the HORIZON-MSCA-2021-SE-01 Research and Innovation Programme under the Marie Sklodowska-Curie grant agreement number 101086388

\end{acknowledgements}

\bibliographystyle{aa}
\bibliography{bib_gaia.bib}

\end{document}